\documentclass[aps,pra,reprint,superscriptaddress,10pt,showpacs,a4paper]{revtex4-1}
\usepackage{amsmath,amssymb,amsfonts}
\usepackage{mathtools}
\usepackage{graphicx}
\usepackage{txfonts}
\usepackage{color, xcolor}

\usepackage[unicode,breaklinks]{hyperref}
\hypersetup{
    unicode=true,
    a4paper=true,
    plainpages=false,
    pdftitle={Predicting Bubble Fragmentation in Superfluids},
    pdfauthor={Jake McMillan, Thomas A. Flynn, Ryan Doran},
    pdfsubject={Predicting Bubble Fragmentation in Superfluids},
    colorlinks=true,
    linkcolor=blue,
    citecolor=blue,
    filecolor=black,
    urlcolor=blue
}
\usepackage{tikz}

\newcommand{\vcrit}{v_{\mathrm{crit}}}

\newcommand{\sech}{\mathrm{sech}}

\newcommand{\WE}{\mathrm{We}_s}

\begin{document}
\title{Predicting Bubble Fragmentation in Superfluids}

\author{Jake McMillan}
\affiliation{Joint Quantum Centre (JQC) Durham--Newcastle, School of Mathematics, Statistics and Physics, Newcastle University, Newcastle upon Tyne, NE1 7RU, UK}

\author{Thomas A. Flynn\address{t.flynn@newcastle.ac.uk}}
\affiliation{Joint Quantum Centre (JQC) Durham--Newcastle, School of Mathematics, Statistics and Physics, Newcastle University, Newcastle upon Tyne, NE1 7RU, UK}

\author{Ryan Doran\address{ryan.doran@newcastle.ac.uk}}
\affiliation{Joint Quantum Centre (JQC) Durham--Newcastle, School of Mathematics, Statistics and Physics, Newcastle University, Newcastle upon Tyne, NE1 7RU, UK}

\date{\today}

\begin{abstract}
In classical fluids, the Weber number is a dimensionless parameter that characterises the flow of a multi-phase fluid. The superfluid analogy of a classical multi-phase fluid can be realised in a system of two or more immiscible Bose-Einstein condensates. These superfluid mixtures have been shown to display a wider variety of exotic dynamics than their single component counterparts. Here we systematically study the dynamics of a binary immiscible Bose-Einstein Condensate in two dimensions, where a small bubble of the second component is used to ``stir'' the first component. We begin by rigorously mapping out the critical velocity for vortex shedding as a function of the size of the bubble, in analogy to the critical velocity of a laser spoon. Observing that the dynamics of the system depend on the initial size and velocity of the bubble, we then show that a dimensionless parameter with the same form as the Weber number accurately predicts the resulting bubble fragmentation.  
\end{abstract}

\maketitle

%% ===================================================================
%% ===================================================================
%% ===================================================================
%% ===================================================================
%% ===================================================================
\section{Introduction}
\label{section:introduction}

From rainfall \cite{Villermaux2009} to liquid jets \cite{Varga2003} and  the smelting of liquid metals \cite{Haas2021}, the dynamics of a wide variety of flows that consist of two or more fluids with an interface can be characterised by the Weber number \cite{Michaelides2017}. For classical multi-phase flows (i.e., droplets of oil contained in water flow), the Weber number is given by
\begin{equation}
    \mathrm{We} = \frac{\rho v^2 l}{\sigma},
    \label{eqn:Weber_number}
\end{equation}
where $\rho$ is the density of the fluid, $v$ is the characteristic velocity of the droplet in the fluid, $l$ is the characteristic size of the droplet, and $\sigma$ is the surface tension at the interface of the droplet. In general, multi-phase flows are non-trivial systems that exhibit complex dynamics. The advantage of the Weber number is that it is a dimensionless parameter based on a small number of observables that broadly characterises the dynamics of the system, without the need for extensive numerical analysis or experimental measurements. In a flow with a very small Weber number, a droplet will retain its original shape. As the Weber number increases in magnitude, vibrational modes will appear on the droplet, leading to deformations and eventually the fragmentation of the droplet. For very large Weber numbers, a catastrophic break-up mode will occur \cite{Guildenbecher2009}, where the original droplet will break into smaller droplets, and these droplets will break into smaller droplets still.

The superfluid analogy of multi-phase fluids have been realised in mixtures of atomic Bose Einstein Condensates (BECs), where the two (or more) components are coupled, and the components may be miscible or immiscible, depending on the inter-species and intra-species interaction strengths \cite{Pu1998}. BEC mixtures can be formed in the same atomic species \cite{Myatt1997,Hall1998,Matthews1999,Miesner1999,Maddaloni2000,Delannoy2001,Schweikhard2004,Mertes2007,Anderson2009,Tojo2010}, different isotopes of the same atomic species \cite{vanKempen2002,Papp2008}, and with different atomic species \cite{Ferrari2002,Modugno2002,Thalhammer2008,McCarron2011}. By comparison with single component BECs, two-component BECs have been shown to exhibit an exotic variety of both steady state solutions \cite{Pu1998,Ho1996,Timmermans1998,Ao1998,Trippenbach2000,Barankov2002,VanSchaeybroeck2008,Guatam2010,Gordon1998,Kim2002}, and dynamics \cite{Li2019,Han2019, Mithun2021,Wheeler2021, Richaud2023}. In addition, recent works on superfluid mixtures have uncovered a rich vein of counterparts to instabilities found in classical fluids. In mixtures that have a mass imbalance, analogies to the Richtmyer-Meshkov \cite{Bezett2010} and Rayleigh-Taylor instabilities have been predicted \cite{Sasaki2009,Kobyakov2014}, while for superfluid systems subject to long-range dipole interactions, ferrofluid instabilities such as the  Rosensweig \cite{Kadau2016} and fingering \cite{Xi2018_fingering} instability have been predicted and experimentally observed. These results hint at the diverse  range of instabilities available in two-component superfluids \cite{Kokubo2021}.

While superfluids are characterised by their frictionless flow about an obstacle, it has been shown that dragging an obstacle through a superfluid faster than a critical velocity will nucleate vortices \cite{Frisch1992}, and increasing the speed of the obstacle further still will lead to a turbulent system \cite{Sasaki2010,Kwon2016}. While there has been much focus on determining the critical velocity as a function of obstacle shape \cite{Stagg2014,Musser2019} and condensate temperature \cite{Stagg2016a,Stagg2016b}, these studies are limited to the case of an external potential acting as an obstacle, which acts as a ``laser spoon'' in creating a well defined region of depleted condensate density. In the absence of an imposed obstacle, it is still possible to create an area of depleted density, by adding an immiscible second component to the system \cite{Hall1998,Miesner1999}, which is subject to arbitrary spatio-temporal control \cite{Gauthier2016}. Previous theoretical work on immiscible binary condensates has observed that a bubble of component 2 will deform and then shed vortices in component 1 as it is subjected to a linear forcing potential \cite{Sasaki2011}. However, it is not possible to identify a critical velocity at which the bubble will shed vortices into the other component, owing to the set-up of the potential. 

In this work, by controlling the imposed velocity profile of the bubble, we systematically study the bubble's critical velocity as a function of its size. The wake of the bubble is determined both by its size and its initial velocity. Small bubbles travelling at low speeds are trailed by laminar flow. As either the bubble's velocity or size is increased, one vortex-antivortex pair is shed, the cores of which are filled by some of the atoms that originally formed the bubble. Large bubbles travelling at high speeds shed many vortices, leaving a dense wake. Most notably, we observe a dynamic similarity between the wakes of bubbles with different sizes and velocities; motivated by previous studies into dynamic similarities \cite{Reeves2015}, we identify a dimensionless quantity that parameterises the resulting dynamics of the system. This quantity, the superfluid Weber number, $\WE$, is based on the classical Weber number, Eqn.~\eqref{eqn:Weber_number}, and is determined by the number of atoms in the bubble and the inter-species interaction strength; two highly controllable experimental parameters. The Weber number also has the advantage that it characterises the dynamics of the system based on the initial configuration, removing the need to disturb the system during its evolution (i.e., to perform Time-of-Flight imaging) to visualise the flow. We show that the value of this number accurately predicts the onset of quantum turbulence, via the irregular shedding of filled vortices. 

The remainder of this letter is structured as follows: in Section~\ref{section:numerics} we introduce the coupled Gross-Pitaevskii equations, and detail our numerical approach. In Section~\ref{section:critical_velocity} we analyse the critical velocity of vortex nucleation that is due to the second component. The main results of this letter are presented in  Section~\ref{section:dynamics}, where we study the dynamics of the system and show that the resulting state can be characterised by a dimensionless parameter in the form of Eqn.~\eqref{eqn:Weber_number}. We present our concluding remarks and outlook in Section~\ref{section:conclusions}.

%% ===================================================================
%% ===================================================================
%% ===================================================================
%% ===================================================================
%% ===================================================================
\section{Governing Equations and Numerical Implementation}
\label{section:numerics}
We consider a binary system of weakly interacting BECs in the zero temperature limit. This consists of a majority component, with macroscopic wavefunction $\psi_1$ and atomic mass $m_1$, which contains a ``bubble'' of a second component, with macroscopic wavefunction $\psi_2$ and atomic mass $m_2$. For simplicity, we consider a homogeneous system with no in-plane trapping potential on either species. Such a system is accurately described by a 2D coupled Gross-Pitaevskii Equation (GPE)
\begin{subequations}
    \begin{align}
    i\hbar\frac{\partial \psi_1}{\partial t} &= \left[ - \frac{\hbar^2}{2m_1} \nabla^2 + u_{11}|\psi_1|^2 + u_{12}|\psi_2|^2 \right]\psi_1, \label{eqn:cgpe1}\\
    i\hbar\frac{\partial \psi_2}{\partial t} &= \left[ - \frac{\hbar^2}{2m_2} \nabla^2 + u_{12}|\psi_1|^2 + u_{22}|\psi_2|^2 \right]\psi_2, \label{eqn:cgpe2}
    \end{align}
\end{subequations}
where sufficient trapping is applied in the $z$ axis to prevent excitations out of the plane \cite{Rooney2011}. We can cast Eqns.~\eqref{eqn:cgpe1} and \eqref{eqn:cgpe2} in dimensionless form by working in the characteristic units of the majority component: background density $n_{1,0}$, healing length $\xi_1 = \hbar/\sqrt{m_1 u_{11} n_{1,0}}$, characteristic time, $\tau = \hbar/\left(u_{11} n_{1,0}\right)$, and speed of sound $c_1=\sqrt{u_{11} n_{1,0}/m_1}$ (see Appendix \ref{app:governing_eqns}). In the remainder of this letter, we will normalise component 1 to $N_1$ and refer to this as the majority component, while component 2 is normalised to $N_2$. The value of $N_1$ is chosen so that the background density of component 1 is approximately unity across our computational domain, $n_{1,0}\approx 1$, and we vary the population of the second component, $N_2$, while fixing the requirement for the density of the bubble to be unity in the bulk, $n_{2,0}\approx 1$  \footnote{This corresponds to a choice of atom number in dimensionless units that is $N_1=L_xL_y - N_2$, where $L_x$ and $L_y$ are the grid sizes in the $x$ and $y$ direction respectively. In order to recover the true atom numbers, this normalisation should be multiplied by $\ell/\left(a_s\sqrt{8\pi}\right)$, where $a_s$ is the s-wave scattering length, and $\ell=\sqrt{\hbar/m \omega_z}$ is the harmonic oscillator length in the $z$ direction associated with a trapping frequency $\omega_z$.}.   We suppose that our binary system comprises a homonuclear systems with equal masses, $m_1=m_2=m$, and equal intra-species interaction parameters, $u_{11}=u_{22}$. This, coupled with the fact that the background density of each species is unity in the bulk, means that the healing length and speed of sound is identical in each component, and as such we will omit the subscripts on the healing length, $\xi$, characteristic time, $\tau$, and speed of sound, $c$.

The formation of a ``bubble'' of the second component is a result of the immiscibility criterion for a homogeneous system \cite{Trippenbach2000}, which constrains the inter-species interaction strength $u_{12}$ as $u_{12}^2>u_{11}u_{22}$. Having set the intra-species scattering lengths to be identical, and taking $u_{12}>0$, we can write this constraint as $g_{12} = u_{12}/\sqrt{u_{11} u_{22}} > 1$, and we will report values of $g_{12}$ in the remainder of this letter.

We solve the coupled GPE, Eqns.~\eqref{eqn:cgpe1} and \eqref{eqn:cgpe2}, using an adaptive RK45 method with a tolerance of $10^{-8}$, implemented using XMDS2 \cite{XMDS2}. We do this on a computational grid that is discretised to have 2 numerical grid points per healing length, typically on a grid of size $256\xi\times128\xi$; where the size of the bubble becomes comparable to the size of this computational domain, we double the linear size. In order to initialise the system, we perform a Wick rotation $t_i=it$ and evolve up to $t_i=100\tau$, re-normalising both components after each step. This ``imaginary time'' propagation is a well established method to obtain the lowest energy state of a system \cite{Barenghi2016}. Once we have obtained the ground state, we impose a phase gradient in the $x$ direction on the bubble, which is responsible for the initial velocity boost, before evolving the system in real time.

%% ===================================================================
%% ===================================================================
%% ===================================================================
%% ===================================================================
%% ===================================================================
\section{The Critical Velocity}
\label{section:critical_velocity}

\begin{figure}
    \centering
    \includegraphics{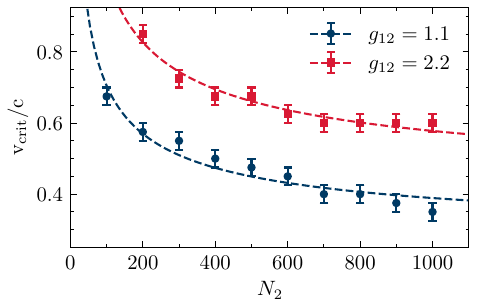}
    \caption{The critical velocity, $v_\mathrm{crit}$ of the bubble containing $N_2$ atoms, where the inter-species interaction strengths are $g_{12}=1.1$ (blue) and $g_{12}=2.2$ (red). The error bars indicate the discrete steps in the initial velocities. Dashed lines indicate the fitted curves given by Eqn.~\eqref{eqn:vc_fit}.}
    \label{fig:critical_velocity}
\end{figure}

%	g=1.1: a = 4.623 +/- 0.449, b = 0.243 +/- 0.024
%   g=2.2: a = 6.376 +/- 0.523, b = 0.376 +/- 0.024

Unlike the case of an obstacle that is imposed via an external potential \cite{Stagg2014,Musser2019,Stagg2016a,Stagg2016b}, a system that is ``stirred'' using an immiscible second component does not have a well defined zero-density region, since the second component can deform from its initial shape. This means that the critical velocity for vortex nucleation in the wake of the bubble can not be predicted analytically using Landau's criterion \cite{Nozieres1999}, and so we determine the critical velocity numerically, as previous works have done for the case of an externally imposed barrier \cite{Stagg2014,Musser2019}. The vortices are quantized due to the superfluid nature of the system.

The results of the vortex detection are presented in Fig.~\ref{fig:critical_velocity}. Since the characteristic diameter of the bubble, $l$, scales as $\sqrt{N_2}$, we would expect that the critical velocity for vortex nucleation, $\vcrit$, will decrease with the number of atoms in the bubble; this is consistent with previous studies for external stirring potentials \cite{Frisch1992,Stagg2014,Reeves2015,Kwon2016}.  Given the qualitative similarities between the immiscible bubble and a stirring potential, we fit an empirical model for the critical velocity, 
\begin{equation}
    \frac{\vcrit}{c} = \frac{a}{\sqrt{N_2}} + b,
    \label{eqn:vc_fit}
\end{equation}
where $b$ corresponds to the ``Eulerian limit'' of very large obstacles \cite{Rica2001}. For $g_{12}=1.1$ we obtain $a=4.623\pm0.449$ and $b=0.243\pm0.024$, while for $g_{12}=2.2$ we obtain $a=6.376\pm0.523$ and $b=0.376\pm0.024$. The form of this fit has previously been used to determine the critical velocity for vortex shedding behind a stirring cylinder with a fixed width \cite{Jopsserand1999,Stagg2014,Stagg2016a}. As $N_2$ increases, the width of the initial bubble increases, and the critical velocity will decrease, asymptotically approaching the corresponding Eulerian limit for a given $N_2$. The effect of increasing $g_{12}$ is to increase the repulsion between the bulk of the majority component and the bubble; this increases the critical velocity of the bubble in analogy with the increase in critical velocity of a hard-walled potential compared to a soft-walled potential \cite{Winiecki1999}.

%% ===================================================================
%% ===================================================================
%% ===================================================================
%% ===================================================================
%% ===================================================================
\section{Characterising the Resulting Dynamics}
\label{section:dynamics}

\begin{figure}
    \centering
    \includegraphics{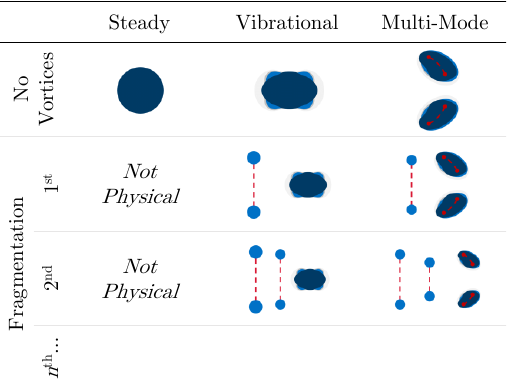}
    \caption{A schematic of the dynamics of a bubble. From states with no vortex shedding, row 1, to states that fragment into 1 vortex-antivortex pair, row 2, to 2 vortex-antivortex pairs, row 3, and so on. Columns indicate the behaviour of the original bubble.}
    \label{fig:break-up_schematic}
\end{figure}

While the critical velocity of a bubble containing $N_2$ atoms of an immiscible second component has a likeness to vortex shedding from a dragged laser spoon, the bubble is not a fixed obstacle, and is able to deform or fragment over the lifetime of the experiment. Studies of two-component classical flows observe a number of ``break-up'' modes of the droplet, termed vibrational, bag, multimode, sheet-thinning, and catastrophic \cite{Guildenbecher2009}. In a system of two immiscible superfluids, we are also able to identify a range of characteristic behaviours, which we categorise based on the long-term behaviour of the fluid parcel that originated from the bubble and the number of vortex pairs that are shed in the wake of this parcel. A schematic of these dynamics can be found in Fig.~\ref{fig:break-up_schematic}.

For very low velocities, we observe a `steady' state where the fluid parcel is relatively unchanged from the original bubble (Fig.~\ref{fig:break-up_schematic}, top left), although any imposed velocity will lead to a ``sloshing'' motion (see the example movies of the dynamics in the supplementary material \cite{Supp}). As the velocity of the bubble is increased, the surface tension at the interface gives rise to vibrational modes in the fluid parcel, similar to the surface modes observed in trapped BECs \cite{Onofrio2000} (Fig.~\ref{fig:break-up_schematic}, top middle). The excitation spectrum of waves at the interface of two BECs has previously been studied \cite{Barankov2002}. As the velocity increases further (or $N_2$ is increased), the vibrational modes of the bubble have sufficient energy that they are able to overcome the surface tension associated with the immiscibility condition, and the bubble breaks into two or more smaller parcels without forming vortices; this is the multi-mode phase (Fig.~\ref{fig:break-up_schematic}, top right). 

As the velocity is increased beyond the critical velocity for a given $N_2$, we begin to observe fragmentation. The response of the superfluid system to this highly non-equilibrium state is to shed an even number of oppositely charged vortices. The $1^\mathrm{st}$ fragmentation occurs when a single vortex-antivortex pair is shed from the leading fluid parcel. The cores of this vortex-antivortex pair are filled with some of the second component, and the leading fluid parcel, which is now smaller than the original bubble, is then either in the vibrational or multi-mode state (Fig.~\ref{fig:break-up_schematic}, second row). 

As the velocity is increased further still, and particularly for larger $N_2$, we observe recursive fragmentation events (see Fig.~\ref{fig:break-up_schematic}, third row). This occurs when the initial fluid bubble fragments by shedding one filled vortex-antivortex pair, but the remaining fluid parcel is still energetically unstable to the shedding of further vortices, and so a second filled vortex-antivortex pair is shed. We refer to this as $2^\mathrm{nd}$ fragmentation. For sufficiently large velocities and $N_2$, we can observe $n^\mathrm{th}$ fragmentation of the initial bubble. On each occasion, a filled vortex-antivortex pair is shed from the leading fluid parcel, reducing the size of the leading fluid parcel, which will either continue to shed filled vortex-antivortex pairs, or it will enter one of the vibrational or multi-mode states. The recursive pattern of vortex shedding and bubble deformation is due to the superfluid nature of the system -- the response of a given bubble to an imposed velocity is to shed vortex-antivortex pairs \cite{Winiecki1999}, due to the fact that vorticity is quantized in a superfluid, and the bubble is depleted after each shedding event, we see a reduced number of dynamical states compared to the classical case.

\begin{figure*}
    \centering
    \includegraphics{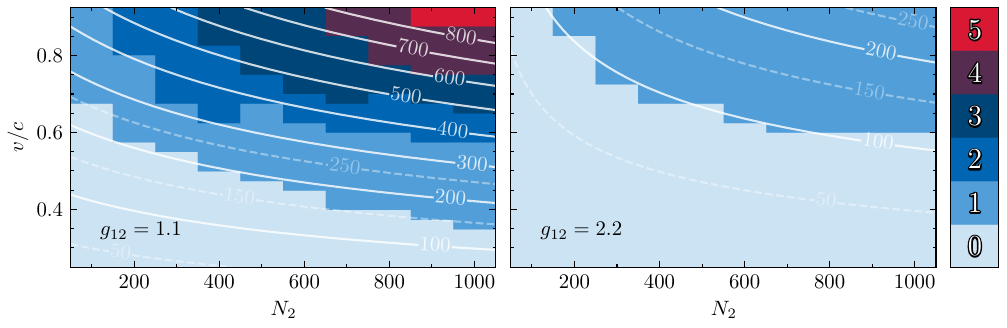}
    \caption{Characterisation of the dynamics for a system with $g_{12}=1.1$, left, and $g_{12}=2.2$, right. The colour axis indicates the observed number of vortex-antivortex pairs that are shed over the lifetime of the simulation, the $n^\mathrm{th}$ fragmentation. The contours are given by the superfluid Weber number, Eqn.~\eqref{eqn:sf_Weber_no}. Example movies of the dynamics are available in the supplementary material \cite{Supp}.}
    \label{fig:characterisation}
\end{figure*}

Having observed that the recursive nature of the bubble dynamics depends on the initial bubble velocity and initial bubble size, it would be useful to be able to predict the bubble dynamics based on some initial parameters. This leads us to a superfluid analog to the classical Weber number, Eqn.~\eqref{eqn:Weber_number}. By approximating the density distribution of the two components to have hyperbolic tanh profiles \cite{Barankov2002}, we can derive the surface tension at the interface of the two fluids to be $\sigma = \xi P_0 \sqrt{g_{12}-1} \tanh\left[\sqrt{N_2\left(g_{12}-1\right)/6\pi}\right]/4\sqrt{6}$, where $P_0=u_{11} n_{1,0}^2/2$ is the pressure (the full derivation can be found in Appendix \ref{app:surf}). We can think of the surface tension as the energy required to deform the density of the components away from a homogeneous profile \cite{VanSchaeybroeck2008}; such a deformation is required by the immiscibility condition of the system, and for a unit area of interface will increase with $g_{12}$. This leads to the  Weber number for a binary system of immiscible superfluids, which is
\begin{equation}
    \WE = \frac{8 \sqrt{6} \rho v^2 \xi}{\sqrt{\pi}} \sqrt{\frac{N_2}{g_{12}-1}} \coth \left[\sqrt{\frac{N_2\left(g_{12}-1\right)}{6\pi}}\right],
    \label{eqn:sf_Weber_no}
\end{equation}
where $\rho$ is the background density of the majority fluid, $v$ is the initial velocity of the bubble, and $l$ is the characteristic size of the bubble, which we approximate as $l=\xi\sqrt{N_2/\pi}$. The precise experimental control over of the interaction parameter, $g_{12}$, and high precision measurements of the atom number, $N_2$, are the hallmark of a superfluid formed of an ultra-cold quantum gas, and we see that these are the most prominent variables in the superfluid Weber number.

Our observed dynamics are presented in Fig.~\ref{fig:characterisation}, along with the predictions of the superfluid Weber number, for each inter-species interaction strength. Although we would expect that the dynamics of any given individual realisation will be subject to small fluctuations about the value predicted by Eqn.~\eqref{eqn:sf_Weber_no}, we see that the contours of the superfluid Weber number are in good agreement with the different regimes of bubble break-up. In particular, we see that $100\lesssim\WE\lesssim 150$ indicates a transition from no fragmentation, to $1^\mathrm{st}$ fragmentation. The presence of vortices in the system associated with fragmentation indicates the breakdown of superfluidity, and the transition of the system to a turbulent state \cite{Winiecki1999,Barenghi2008,Moon2015}. In the case where $g_{12}=1.1$, we see a wide range of dynamics, from no fragmentation (light blue), up to $5^\mathrm{th}$ fragmentation, which has a superfluid Weber number $\WE\gtrsim 800$. The dynamics of the system for this large superfluid Weber number is characterised by a large number of filled vortices, which is an inherently chaotic system that contains large velocity gradients; as $\WE$ is increased further, these systems will become turbulent, as has been seen experimentally in one component \cite{Kwon2016}. This transition to turbulence takes place when the dimensionless Weber number is an order of magnitude greater than the Weber number associated with $1^\mathrm{st}$ fragmentation. We see the best agreement between the prediction of the superfluid Weber number and the ensuing dynamics of the system occurs at larger $N_2$, this is because the approximation of the density profile of the two components is more accurate in this regime.  In the case where $g_{12}=2.2$, we are limited to $1^\mathrm{st}$ fragmentation for the values of $N_2$ that we consider -- this is due to the stronger immiscibility condition associated with the higher $g_{12}$ preventing the initial bubble from breaking up, other than at high velocity or large bubble size. The superfluid Weber number correctly predicts that this will be the case.

%% ===================================================================
%% ===================================================================
%% ===================================================================
%% ===================================================================
%% ===================================================================
\section{Conclusions}
\label{section:conclusions}

For a binary system of immiscible superfluids, we have systematically determined the critical velocity for vortex shedding by a bubble of the second component. Such a system would be relatively easy to realise using current experimental set-ups, for example by making use of DMDs, \cite{Gauthier2016}. We have then studied the resulting dynamics, and we have observed that they can be characterised by the number of vortices that are shed, and the behaviour of the fluid parcel that remains from the bubble. Importantly, since vorticity is quantized, the dynamics are recursive -- the bubble undergoes $n$ fragmentation events (shedding $n$ vortex-antivortex pairs) before reaching a vibrational or multi-mode state. The dynamics of the bubble wake depend on the initial velocity of the bubble, and the size of the bubble; we have shown that this wake can be characterised by a dimensionless parameter, the superfluid Weber number $\WE$, which resembles the form of a Weber number in classical multi-phase flows, Eqn.~\eqref{eqn:Weber_number}. Like the classical Weber number, the superfluid Weber number parameterises the dynamics of the system, and is a useful starting point to predict the nature of the resulting flow, with steady or vortex-free flow taking place when $\WE\ll100$, the shedding of one vortex-antivortex pair occurring at $100\lesssim\WE\lesssim 150$, and $5^\mathrm{th}$ fragmentation state occurs at $\WE\gtrsim 800$. Such dynamic similarities suggest that the superfluid Weber number may be applicable to a wide variety of superfluid systems, from binary immiscible Bose-Einstein condensates, to systems containing multiple phases of liquid helium. Excitingly, we have seen that the superfluid Weber number is particularly good when $N_2$ is large or $g_{12}$ is strong, the regime predicted to exhibit new forms of superfluid turbulence that are unavailable in one-component systems.

%% ===================================================================
%% ===================================================================
%% ===================================================================
%% ===================================================================
%% ===================================================================

\acknowledgements 
We thank Prof. Nick Parker for helpful discussions during the preparation of this manuscript.  This work made use of the Rocket HPC facility at Newcastle University. JM thanks Newcastle University for a Vacation Research Scholarship. TAF acknowledges support from the UK Engineering and Physical Sciences Research Council, Grant No. EP/T517914/1;  RD is supported by the UK Engineering and Physical Sciences Research Council, Grant No. EP/X028518/1.

%% ===================================================================
%% ===================================================================
%% ===================================================================
%% ===================================================================
%% ===================================================================
\appendix
\section{Governing Equations}
\label{app:governing_eqns}
A binary system of weakly-interacting BECs in the zero-temperature limit is accurately described by the coupled Gross-Pitaevskii Equation (GPE)
\begin{subequations} 
\begin{align}
    i\hbar \frac{\partial \Psi_1}{\partial t} &=& \left[ - \frac{\hbar^2}{2m_1} \nabla^2 + V_1 + U_{11} |\Psi_1|^2 + U_{12} |\Psi_2|^2\right] \Psi_1,  \\
    i\hbar \frac{\partial \Psi_2}{\partial t} &=& \left[ - \frac{\hbar^2}{2m_2} \nabla^2 + V_2  + U_{12} |\Psi_1|^2 + U_{22} |\Psi_2|^2\right] \Psi_2,    
\end{align}
\end{subequations}
where $\Psi_k$, $m_k$, and $V_k$ are the macroscopic wavefunction, the mass of the atomic species, and the external trapping potential of the $k$-th components, respectively. The parameters $U_{11}$ and $U_{22}$ represent the \emph{intra}-species interaction strengths of components 1 and 2 respectively,
\begin{equation}
    U_{jj} = \frac{4\pi\hbar^2 a_{jj}}{m_j},
\end{equation}
where $j\in\{1,2\}$ and $a_{jj}$ is the scattering length between atoms in component $j$. The parameter $U_{12}$ represents the \emph{inter}-species interaction between component 1 and 2,
\begin{equation}
    U_{12} = \frac{2\pi\hbar^2 a_{12}}{m_\mathrm{red}},
\end{equation}
where $a_{12}$ is the scattering length between atoms in different components and $m_\mathrm{red}^{-1}=m_1^{-1}+m_2^{-1}$ is the reduced mass. For a homogeneous system, the immiscibility condition is \cite{Trippenbach2000}
\begin{equation}
    U_{12}^2 > U_{11} U_{22}. 
\end{equation}

In order to study the dynamics of a quasi-2D system, we take $V_j=m_j \omega_j^2 z^2 /2$ for $j=1,2$, where the trapping frequency $\omega_j$ is sufficiently strong that the excitation of modes is prevented in the $z$ direction. This allows us to write the $j$-th wavefunction as $\Psi_j = \psi_j(x,y,t) \exp\left(-z^2/2\ell_j^2\right)$, where $\ell_j=\sqrt{\hbar/m_j\omega_j}$ is the harmonic oscillator length and $\psi_j$ is the wavefunction in the $xy$ plane. We assume that our binary system comprises a homonuclear mixture of condensates with identical masses ($m=m_1=m_2$) and intra-species scattering length $a_\mathrm{s}$, subject to identical trapping frequencies. This means that the harmonic oscillator lengths are identical ($\ell=\ell_1=\ell_2$), and the effective interaction parameters are 
\begin{equation}
    u_{jk} = \frac{U_{jk}}{\sqrt{2\pi}\ell},
\end{equation}
for $j,k\in\{1,2\}$.

Without loss of generality, we will consider component 1 to be the main component. Aside from the strong harmonic trapping that forces the system to be quasi-2D, there are no external trapping potentials in the system, and so the ground state of component 1 has a uniform background density, $n_{1,0}$. It is useful to work in the natural units of component 1, so length is cast in terms of the healing length
\begin{equation}
\xi=\frac{\hbar}{\sqrt{m u_{11} n_{1,0}}},
    \label{eqn_supp:healing_length}
\end{equation}
the characteristic energy of the system is given by $n_{1,0} u_{11}$, and time is given by $\tau=\hbar/\left(n_{1,0}u_{11}\right)$. This means that the characteristic speed of the system is the speed of sound, $c=\xi/\tau$. Working in these units, the governing equations of the 2D system can be written in non-dimensionalised form as
\begin{subequations}
    \begin{align}
        i \frac{\partial \psi_1}{\partial t} &= \left(-\frac{1}{2}\nabla_\perp^2 + |\psi_1|^2 + g_{12}|\psi_2|^2\right)\psi_1, \label{eqn:app_cgpe1} \\
        i \frac{\partial \psi_2}{\partial t} &= \left(-\frac{1}{2}\nabla_\perp^2 + |\psi_2|^2 + g_{12}|\psi_1|^2\right)\psi_2, \label{eqn:app_cgpe2}
    \end{align}
\end{subequations}
where the dimensionless interaction parameter 
\begin{equation}
    g_{12} = \frac{u_{12}}{\sqrt{u_{11}u_{22}}}.
    \label{eqn_supp:dimensionless_interactions}
\end{equation}
In these units, the immiscibility criterion for a homogeneous system \cite{Trippenbach2000} reduces to $g_{12}>1$. In the remainder of this paper, we will set the normalisation of components 1 and 2 to be $N_1$ and $N_2$ respectively, however, we note that these numbers should be multiplied by $\ell/\left(\sqrt{8\pi}a_\mathrm{s}\right)$ to recover the true atom number. 

We solve the coupled GPE, Eqns.~\eqref{eqn:app_cgpe1} and \eqref{eqn:app_cgpe2} using an adaptive RK45 method with a tolerance of $10^{-8}$, implemented using XMDS2 \cite{XMDS2}. We do this in on a computational grid that is discretised to have 2 numerical grid points per healing length, typically on a grid of size $256\xi\times128\xi$. Where the size of the bubble becomes comparable to the size of this computational domain, we double the linear size. In order to initialise the system, we perform a Wick rotation $t_i=it$ and evolve up to $t_i=100\tau$, re-normalising both components after each step. This ``imaginary time'' propagation is a well established method to obtain the lowest energy state of a system \cite{Barenghi2016}. Once we have obtained the ground state we impose the phase 
\begin{equation}
    \psi_2 \longrightarrow \psi_2 \exp \left( - \frac{2\pi i v_{int} x}{L_x} \right),
\end{equation}
 where $L_x$ is the length of the computational domain in the $x$ direction and $v_{int}$ is the nearest integer to $vL_x/2\pi$, which is responsible for the initial velocity boost on the second component, before evolving the system in real time.

\section{Surface Tension in a 2D bubble}
\label{app:surf}
\begin{figure}
    \centering    \includegraphics{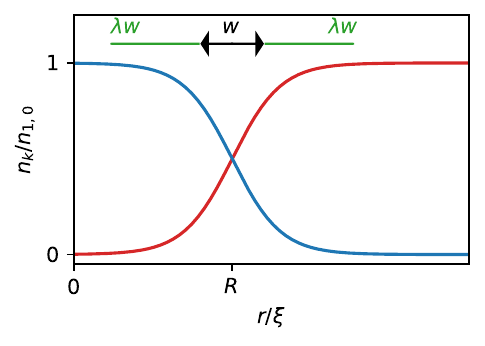}
    \caption{A radial slice of the density given in the ansatz, Eqn.~\eqref{supp_eqn:ansatz1} in red  and Eqn.~\eqref{supp_eqn:ansatz2} in blue, for $N_2=2000$ and $g_{12}=1.1$. The interface width, obtained in Eqn.~\eqref{eqn_supp:width}, is shown as a guide to the eye, the radius of the interface $R=\sqrt{N_2/\pi}$.}
\label{fig:supp_interface_ansatz}
\end{figure}

%\subsection{The Surface Tension}
Suppose that we are in a 2D system, with the centre of the bubble of component 2 at the origin, and the bubble is axially symmetric. Working in polar coordinates $\left(r,\theta\right)$, with $0\leq r$ and $\theta\in[0,2\pi)$, we can approximate 
\begin{subequations}
    \begin{align}
    n_1(r) &= \frac{1}{2} n_{1,0}\left[1+\tanh\left(\frac{r-R}{w}\right)\right], \label{supp_eqn:ansatz1} \\
     n_2(r) &= \frac{1}{2} n_{2,0}\left[1-\tanh\left(\frac{r-R}{w}\right)\right],
     \label{supp_eqn:ansatz2}
    \end{align}
    \end{subequations}
where $R$ is the radius of the interface, and $w$ encodes the width of the interface. A schematic of these density profiles can be found in Fig.~\ref{fig:supp_interface_ansatz}. In this system, the total surface tension of a bubble can be calculated by 
\begin{equation}
    \Sigma = \frac{1}{2} \int_0^{2\pi}\ d\theta \ \int_0^\infty \ r \, dr \ \sum_{j=1,2} \frac{\hbar^2}{2m_j} \left(\frac{d\sqrt{n_j}}{dr}\right)^2,
\end{equation}
where the $n_j$ are a function of $r$ only. We note that the derivative term can be written as 
\begin{equation}
    \left(\frac{d\sqrt{n_j}}{dr}\right)^2 = \left( \frac{1}{2} \frac{1}{\sqrt{n_j}} \frac{dn_j}{dr}\right)^2 = \frac{1}{4} \frac{1}{n_j} \left(\frac{dn_j}{dr}\right)^2.
\end{equation}

In what remains, we need to calculate 
\begin{equation}
    \Sigma = \frac{2\pi}{2} \int_0^\infty \ r \, dr \ \left[\frac{\hbar^2}{2m_1} \frac{1}{4n_1}\left(\frac{dn_1}{dr}\right)^2 + \frac{\hbar^2}{2m_2}\frac{1}{4n_2}\left(\frac{dn_2}{dr}\right)^2\right],
\end{equation}
which, if we assume that $m_1=m_2=m$, is 
\begin{equation}
    \Sigma = \frac{\pi\hbar^2}{8m}\int_0^\infty \ r \, dr \ \left[ \frac{1}{n_1}\left(\frac{dn_1}{dr}\right)^2 + \frac{1}{n_2}\left(\frac{dn_2}{dr}\right)^2\right].
\end{equation}

To begin, we can calculate the derivatives, 
\begin{subequations}
    \begin{align}
        \frac{dn_1}{dr} &= \frac{1}{2}\frac{n_{1,0}}{w}\sech^2\left(\frac{r-R}{w}\right), \\
        \frac{dn_2}{dr} &= - \frac{1}{2}\frac{n_{2,0}}{w}\sech^2\left(\frac{r-R}{w}\right).
    \end{align}
\end{subequations}
We can now calculate the terms within the square brackets in the integrand, which are 
\begin{eqnarray}
     %\frac{1}{n_1}\left(\frac{dn_1}{dr}\right)^2 &+& \frac{1}{n_2}\left(\frac{dn_2}{dr}\right)^2 \nonumber \\ 
    %&=&
    \sum_{j=1,2} \frac{1}{n_j}\left(\frac{dn_j}{dr}\right)^2 = \frac{1}{2 w^2}\left[ \frac{n_{1,0}\sech^4\left(\frac{r-R}{w}\right)}{1+\tanh\left(\frac{r-R}{w}\right)} + \frac{n_{2,0}\sech^4\left(\frac{r-R}{w}\right)}{1-\tanh\left(\frac{r-R}{w}\right)}\right]. \nonumber \\
\end{eqnarray}
Using the hyperbolic form of the Pythagorean identity we can take out a global factor of $\textrm{sech}^2\left(\frac{r-R}{w}\right)$, and then write
\begin{eqnarray}
    \textrm{sech}^2\left(\frac{r-R}{w}\right)  
    = \left[ 1+ \tanh\left(\frac{r-R}{w}\right)\right]\left[1-\tanh\left(\frac{r-R}{w}\right)\right]. \nonumber \\
\end{eqnarray}
This allows us to cancel the denominators within the square brackets in order to get
\begin{eqnarray}
    \Sigma &=& \frac{\pi \hbar^2}{16mw^2}\left(n_{1,0}+n_{2,0}\right)\int_0^\infty \ dr \ r\sech^2\left(\frac{r-R}{2}\right) \nonumber \\
    &+& \frac{\pi \hbar^2}{16mw^2}\left(n_{2,0}-n_{1,0}\right) \int_0^\infty \ dr \ r \sech^2\left(\frac{r-R}{w}\right)\tanh\left(\frac{r-R}{w}\right). \nonumber \\
\end{eqnarray}
Now, if we assume that $n_{1,0}=n_{2,0}$, we only need to calculate the first integral.

We now write the result
\begin{equation}
    \int r \sech^2 \left(\frac{r-R}{w}\right) dr = w r \tanh \left(\frac{r-R}{w}\right) - w^2 \ln \left[ \cosh \left(\frac{r-R}{w}\right)\right],
\end{equation}
up to a constant. Away from the boundary, the derivative of $n_1$ and $n_2$ is approximately zero. We therefore calculate the integral over the region $r\in\left(R-\lambda w, R+\lambda w\right)$, where $\lambda$ is a dimensionless positive real number [see Fig.~\ref{fig:supp_interface_ansatz}]. Under this approximation, the total surface tension is  
\begin{equation}
    \Sigma = \frac{\pi \hbar^2 n_{1,0}}{8mw} \left\{r\tanh\left(\frac{r-R}{w}\right) - w\ln\left[\cosh\left(\frac{r-R}{w}\right)\right]\right\}^{R+\lambda w}_{R-\lambda w},
\end{equation}
which leads to 
\begin{equation}
    \Sigma = \frac{\pi \hbar^2 n_{1,0} R}{4mw}\tanh \lambda. 
\end{equation}

We use Eqn.~\eqref{eqn_supp:healing_length} for the healing length, and take 
\begin{equation}
    P_0 = \frac{1}{2} u_{11} n_{1,0}^2
    \label{eqn_supp:pressure}
\end{equation}
for background pressure. If $\Sigma$ is the total surface tension (or the total excess energy density at the interface of the bubble), then the surface tension per unit length is given by 
\begin{equation}
    \sigma = \frac{\Sigma}{2\pi R}.
    \label{eqn_sup:st}
\end{equation}
For clarity, we can write $R=R'\xi$ and $w=w'\xi$, where $R'$ and $w'$ are dimensionless quantities. If we substitute this into Eqn.~\eqref{eqn_sup:st} and use the expressions Eqn.~\eqref{eqn_supp:healing_length} for $\xi$ and Eqn.~\eqref{eqn_supp:pressure} for $P_0$ then we have 
\begin{equation}
    \sigma = \frac{\xi P_0}{4w'} \tanh \lambda.
    \label{eqn_supp:sigma_lambda}
\end{equation}

Working with effective 2D interaction parameters, Timmerman's expression for the boundary depth \cite{Timmermans1998} can be written as
\begin{equation}
    w = \sqrt{3} \sqrt{\frac{\xi_1^2 + \xi_2^2}{u_{12}/\sqrt{u_{11} u_{22}} -1}},
\end{equation}
where $\xi_j$ is the healing length of the $j$-th component. Under the assumption that the masses and intra-species interaction strengths of the two species are identical, this reduces to 
\begin{equation}
    w = \frac{\sqrt{6}}{\sqrt{g_{12} - 1}} \xi,
    \label{eqn_supp:width}
\end{equation}
where $g_{12}$ is the dimensionless interaction parameter introduced in Eqn.~\eqref{eqn_supp:dimensionless_interactions}. By comparison with Eqn.~\eqref{eqn_supp:sigma_lambda}, this gives us an estimate of the dimensionless interface width $w'=\sqrt{6}/\sqrt{g_{12}-1}$.

%\subsection{The choice of the parameter $\lambda$.}
We now discuss the choice of $\lambda$, since it appears in Eqn.~\eqref{eqn_supp:sigma_lambda}. We wish to choose $\lambda$ to be large enough that it contains all of the necessary contributions from the density changes at the edge of the bubble, whilst being small enough that $R'-\lambda w' \geq 0$. This final inequality depends on the atom number, since $R\approx\sqrt{N_2/\pi}$ for large $N_2$, and the interaction strength $g_{12}$, since $w'\to 0$ as $g_{12}\to \infty$, and $w'\to\infty$ as $g_{12}\to 0$. If we take $R'\approx\sqrt{N_2/\pi}$ and use the expression for $w'$ from above, this inequality can be written as
\begin{equation}
    \sqrt{\frac{N_2}{\pi}} - \lambda \sqrt{\frac{6}{g_{12}-1}} \geq 0.
\end{equation}
Rearranging this, we find that the maximum value of the parameter $\lambda$ is 
\begin{equation}
    \lambda_\mathrm{max} = \sqrt{\frac{N_2\left(g_{12} -1\right)}{6\pi}},
\end{equation}
and we take this to be the value of $\lambda$ in the remainder of our analysis. Substituting these results into Eqn.~\eqref{eqn_supp:sigma_lambda} we obtain 
\begin{equation}
    \sigma = \frac{1}{4\sqrt{6}} \xi P_0 \sqrt{g_{12}-1} \tanh \left[\sqrt{\frac{N_2\left(g_{12}-1\right)}{6\pi}}\right].
\end{equation}

\section{Forming the Weber number}
\label{app:web}
The form of the classical Weber number is $\mathrm{We}=\rho v^2 l/\sigma$, Eqn.~(1), where the symbols are described in the main text. We can write density as $\rho = \rho' m n_{1,0}$ and the velocity as  $v = v'\xi / \tau $, where $\rho'$ and $v'$ are dimensionless parameters. We approximate the dimensionless radius of the bubble to be $\sqrt{N_2/\pi}$, so that $l=\xi\sqrt{N_2/\pi}$. In the previous section, we have derived an expression for the surface tension per unit length, and so we can substitute these quantities into the classical form of the Weber number to get
\begin{eqnarray}
    \mathrm{We} = \frac{8\sqrt{6}}{\sqrt{\pi}} \sqrt{\frac{N_2}{g_{12}-1}} \rho' \left(v'\right)^2\coth\left[\sqrt{\frac{N_2\left(g_{12}-1\right)}{6\pi}}\right]. \nonumber \\ 
\end{eqnarray}

%% ===================================================================
%% ===================================================================
%% ===================================================================
%% ===================================================================
%% ===================================================================
\bibliography{Weber_no_references.bib}

\end{document}